\documentclass[3p,times,procedia]{elsarticle}
\usepackage{nupha_ecrc}
\usepackage{color}

\volume{00}

\firstpage{1}

\journalname{Nuclear Physics A}

\runauth{K. Murase et al.}

\jid{nupha}

\jnltitlelogo{Nuclear Physics A}

\usepackage{amssymb,amsmath,bm}

\usepackage[figuresright]{rotating}

\begin{document}

\begin{frontmatter}



\dochead{}

\title{Hydrodynamic fluctuations and dissipation in an integrated dynamical model}


\author[a,b,c]{Koichi Murase}
\author[a]{Tetsufumi Hirano}
\address[a]{Department of Physics, Sophia University, Tokyo 102-8554, Japan}
\address[b]{Department of Physics, the University of Tokyo 113-0033, Japan}
\address[c]{Theoretical Research Division, Nishina Center, Riken, Wako 351-0198, Japan}

\begin{abstract}
We develop a new integrated dynamical model
to investigate the effects of the hydrodynamic fluctuations
on observables in high-energy nuclear collisions.
We implement hydrodynamic fluctuations
in a fully 3-D dynamical model consisting
of the hydrodynamic initialization models of the Monte-Carlo Kharzeev-Levin-Nardi model,
causal dissipative hydrodynamics and the subsequent hadronic cascades.
By analyzing the hadron distributions obtained by massive event-by-event simulations
with both of hydrodynamic fluctuations and initial-state fluctuations,
we discuss the effects of hydrodynamic fluctuations on the flow harmonics,
$v_n$ and their fluctuations.
\end{abstract}


\end{frontmatter}


\section{Introduction}\label{}
In high-energy nuclear collisions,
initial-state fluctuations are the major origin
of event-by-event fluctuations observed
in the higher-order azimuthal anisotropies $v_n$ of the observed hadrons.
Since the hydrodynamic response of the matter
converts the initial spatial anisotropies to the final momentum anisotropies,
we can extract transport properties of the created matter, such as viscosity and relaxation times,
by comparing the initial-state fluctuations and the final observables.
However the initial-state fluctuations are not the only source of the event-by-event fluctuations
\cite{Kapusta:2011gt}.
For example, hydrodynamic fluctuations
generate additional flow fluctuations
during the hydrodynamic evolution of the matter.
Hydrodynamic fluctuations are
the thermal fluctuations of the hydrodynamic fields
whose power is related to transport coefficients
through the fluctuation-dissipation relation (FDR)
\cite{Landau:1980mil}.
To precisely determine the transport properties of the matter,
we need to quantify the effects from the hydrodynamic fluctuations
on the observables such as the flow coefficients $v_n$.
We implement hydrodynamic fluctuations
into our dynamical model of high-energy nuclear collisions
to study the effects.

\def\ud#1#2{^{#1}{}_{#2}}
\section{Integrated dynamical model with hydrodynamic fluctuations}
We use an extended version
of an integrated dynamical model
\cite{Hirano:2012kj}
consisting of four parts.
First, event-by-event initial conditions
are generated using the Monte-Carlo Kharzeev-Levin-Nardi (MC-KLN) model
\cite{Drescher:2007ax}.
Subsequent hydrodynamic expansion is calculated using newly developed codes
of the second-order relativistic fluctuating hydrodynamics,
namely viscous hydrodynamics with hydrodynamic fluctuations.
Then we sample hadrons
on the isothermal hypersurface at the temperature of 155 MeV
using the Cooper-Frye formula
\cite{Cooper:1974mv}
including the viscous correction of phase-space distribution $\delta f$
\cite{Teaney:2003kp}.
Finally we perform hadronic cascades using JAM
\cite{Nara:1999dz} to obtain hadron spectra.

In our hydrodynamic calculations
the time evolution of the created matter
is solved in the $\tau$-$\eta_s$ coordinates:
($\tau$, $\eta_s$, $\bm{x}_\perp$) where $t=\tau\cosh\eta_s$ and $z=\tau\sinh\eta_s$.
An essential part of hydrodynamic equations
is the conservation law of the energy-momentum tensor:
\begin{align}
  \bar\partial_\mu T^{\mu\nu} &= 0, &
  T^{\mu\nu} &= eu^\mu u^\nu - P\Delta^{\mu\nu} + \pi^{\mu\nu},
\end{align}
where $\bar\partial_\mu$ is the covariant derivative,
and $e$, $P$ and $\pi^{\mu\nu}$ are
the energy density, the equilibrium pressure and the shear stress,
respectively.
The bulk pressure is not considered here.
The Landau frame is adopted for
a fluid velocity $u^\mu$ ($u_\mu u^\mu = 1$),
where $T^\mu{}_\nu u^\nu = eu^\mu$.
Here the signs of metric tensor $g_{\mu\nu}$ are $(+,-,-,-)$.
The tensor $\Delta^{\mu\nu} = g^{\mu\nu} - u^\mu u^\nu$ is the projector
onto the space of four-vectors transverse to the flow velocity.
For the equation of state (EoS), $P(e)$, we adopt $s95p$-v1.1
\cite{Huovinen:2009yb}
which smoothly connects a lattice EoS at high temperature
and a hadron gas EoS at low temperature.
For the evolution of the shear stress,
the following constitutive equation is used:
\begin{align}
  \tau_\pi \Delta^{\mu\nu}{}_{\alpha\beta} u^\lambda \bar\partial_\lambda \pi^{\alpha\beta}
  + \pi^{\mu\nu} [ 1+ (4/3)\tau_\pi \bar\partial_\lambda u^\lambda]
  &= 2\eta\Delta^{\mu\nu}{}_{\alpha\beta} \bar\partial^{\alpha} u^\beta + \xi^{\mu\nu},
\end{align}
where $\Delta^{\mu\nu}{}_{\alpha\beta}
= \frac12(\Delta\ud\mu\alpha\Delta\ud\nu\beta + \Delta\ud\mu\beta\Delta\ud\nu\alpha)
-\frac13\Delta^{\mu\nu}\Delta_{\alpha\beta}$ is the projector
onto the tensor components
which are symmetric, traceless and transverse to the flow velocity.
In this study,
the shear viscosity coefficient $\eta$
is fixed to be the KSS bound: $\eta/s = 1/4\pi$
\cite{Kovtun:2004de}.
The relaxation time is chosen as $\tau_\pi = 3/4\pi T$
\cite{Song:2009gc,Baier:2007ix}.
Hydrodynamic fluctuations appear
as the term $\xi^{\mu\nu}$
which is the Gaussian white noise
with autocorrelations determined by the FDR
\cite{Murase:2013tma}:
\begin{align}
  \langle \xi^{\mu\nu}(x)\xi^{\alpha\beta}(x')\rangle
    &= 4T\eta \Delta^{\mu\nu\alpha\beta} \delta^{(4)}(x-x').
\end{align}
Noise values are numerically generated using Gaussian random numbers.
The above autocorrelations are diagonalized
by considering proper linear combinations of noise components: $\xi^{(i)}$ ($i=1,\dots,5$),
so that correlations among different components are disentangled.
Then independent noise components $\xi^{(i)}$ are generated as:
\begin{align}
  \xi^{(i)}(\tau,\eta_s,\bm{x}_\perp) &= \sqrt{4T\eta\Delta^{(i)}/\tau}\, w_i(\tau,\eta_s,\bm{x}_\perp),
  \label{eq:gen-noise}
\end{align}
where $w_i(x)$ are the normal Gaussian noise fields,
and $\Delta^{(i)}$ are coefficients coming from the projector.
The Jacobian $1/\tau = 1/\sqrt{-g}$ for the $\tau$-$\eta_s$ coordinates
comes from a coordinate transform of the delta function.

Here we introduce a coarse-graining scale
or a cutoff length scale of the hydrodynamic fluctuations.
The hydrodynamic description of physical systems
has always a lower-bound scale of the description
where the assumption of the local equilibrium becomes invalid.
If hydrodynamic fluctuations
with arbitrarily small wavelength are considered,
the magnitude of fluctuations diverges
such that the stress tensor becomes unphysically large
and breaks calculations.
Therefore we introduce a coarse-graining scale of the hydrodynamic fluctuations
by replacing the normal noise fields $w_i(x)$ in \eqref{eq:gen-noise} with smeared ones $w'_i(x)$:
\begin{align}
  w'_i(\tau,\eta_s,\bm{x}_\perp) &= \int d\eta_s' d^2x'_\perp
    \frac1{\sqrt{(2\pi)^3}\sigma_\eta\sigma_\perp^2}
    \exp\left(- \frac{(\eta_s-\eta_s')^2}{2\sigma_\eta^2} -\frac{(\bm{x}_\perp-\bm{x}'_\perp)^2}{2\sigma_\perp^2}\right)
    w_i(\tau,\eta_s',\bm{x}'_\perp).
\end{align}
For this study, we choose smearing scales
as $\sigma_\perp=1\;\textrm{fm}$, and $\sigma_\eta=1$.

\section{Results}
To quantify the effects of hydrodynamic fluctuations on observables,
we perform event-by-event simulations using the integrated dynamical model.
For the collision system
we consider minimum-bias Au+Au collisions at the collision energy $\sqrt{s_\mathrm{NN}}=$ 200 GeV.
We consider two types of hydrodynamic calculations:
fluctuating hydrodynamics with the shear viscosity and the corresponding hydrodynamic fluctuations,
and conventional viscous hydrodynamics without hydrodynamic fluctuations.
The initial-state entropy densities
are scaled for each type of hydrodynamics
to reproduce experimental charged particle multiplicities
\cite{Adler:2003cb}.
For each type of hydrodynamics
we perform $10^5$ events of hydrodynamic simulations.
Then one hundred events of hadronic cascades are performed
for each hydrodynamic event,
i.e., $10^5\times 100 = 10^7$ cascades
are performed in total for each type of hydrodynamics.

\begin{figure}[!hbt]
  \centering
  \includegraphics[width=0.40\textwidth]{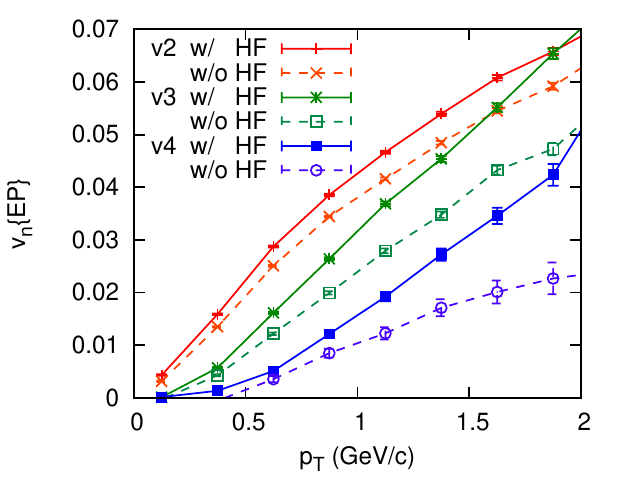}\quad\quad
  \includegraphics[width=0.40\textwidth]{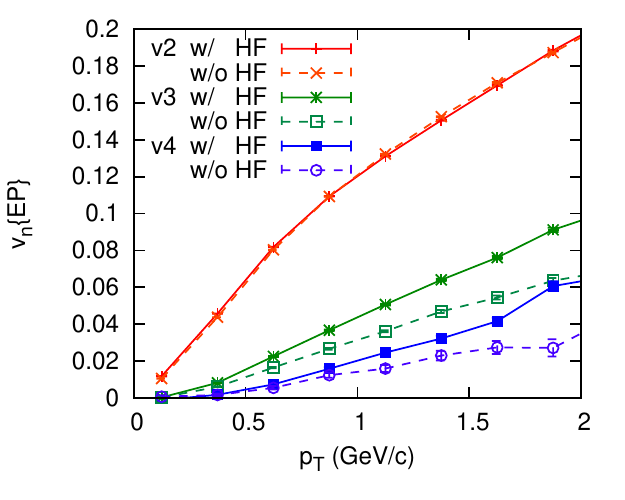}
  \caption{The $p_T$ differential flows
  of the charged hadrons are shown
  for central collisions 0-5\% (the left panel),
  and for non-central collisions 20-30\% (the right panel).
  The flow coefficients are calculated using the $\eta$-sub method
  with two subevents at the pseudorapidity range $1 < |\eta| < 2.8$.
  The solid lines are the results from fluctuating hydrodynamics
  with hydrodynamic fluctuations (HF),
  and the dashed ones are from conventional viscous hydrodynamics.}\label{fig:vnEP}
\end{figure}
In the left panel of Fig.~\ref{fig:vnEP},
we see increase of the flow harmonics $v_n$
in central collisions 0-5\%
due to hydrodynamic fluctuations.
The amount of increase is larger
for higher order of the harmonics,
which can be understood by the FDR:
The effective size of the fluctuations
scales as $\xi \propto 1/\sqrt{V}$
since the delta function in the FDR becomes $1/V$
through effective coarse graining in the volume $V$.
As a higher order of harmonics corresponds to
smaller structures of the created matter,
the effects are larger for the higher order.

\begin{figure}[!hbt]
  \centering
  \includegraphics[width=0.32\textwidth]{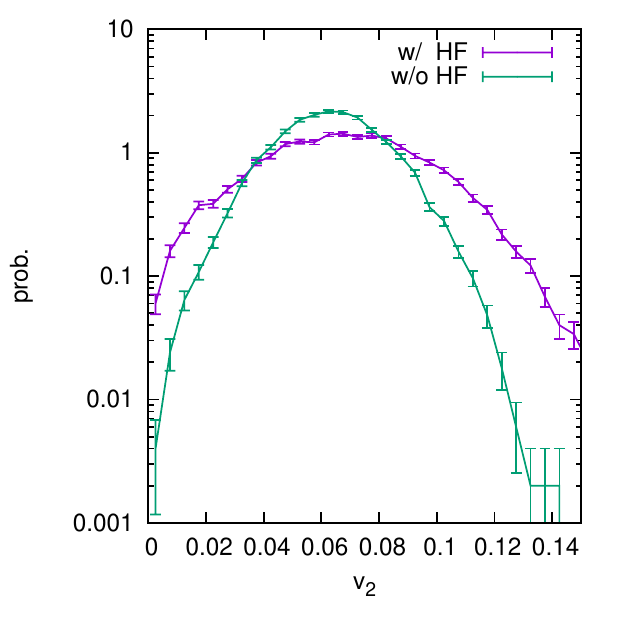}\quad\quad
  \includegraphics[width=0.40\textwidth]{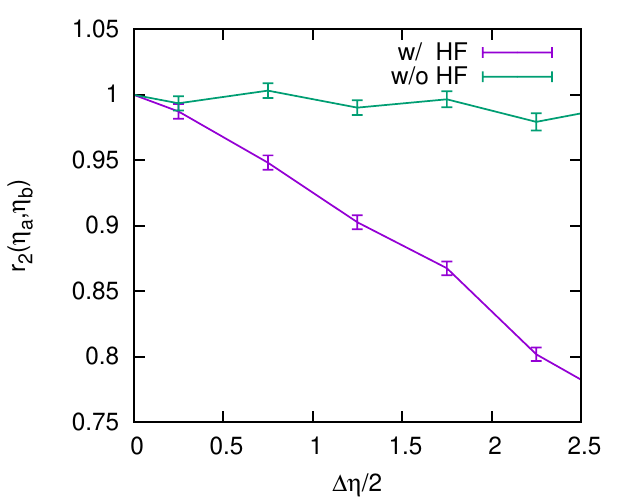}
  \caption{In the left panel, the event-by-event distributions
  of the $p_T$-integrated $v_2$
  for non-central collisions 20-30\% are shown.
  The vertical axis indicates the probability density of $v_n$.
  In the right panel, $\eta$-dependent factorization ratio $r_2(\eta_a,\eta_b)$
  proposed in Ref.~\cite{Khachatryan:2015oea}
  is shown for non-central collisions.
  The horizontal axis is $\eta_a = \Delta\eta/2$.
  The rapidity range for $\eta_b$ is fixed as $3 < \eta_b < 3.5$.}
  \label{fig:vnExE}
\end{figure}

The right panel of Fig.~\ref{fig:vnEP}
shows the same results for non-central collisions 20-30\%.
Unlike the case of the central collisions,
the elliptic flows $v_2$ are almost identical with each other in both cases,
which can be explained by a difference of the origin of elliptic flows.
In central collisions,
the origin of elliptic flows
is dominated by fluctuations.
In non-central collisions,
there is an additional origin from the collision geometry.
Small flow fluctuations caused by hydrodynamic fluctuations
around the large geometrical elliptic flow
do not change the average magnitude of the flows.
Nevertheless the distribution of the flows
is changed by hydrodynamic fluctuations.
This can be confirmed in Fig.~\ref{fig:vnExE}
in which the event-by-event distribution
of $v_2$ in the non-central collisions is broadened by hydrodynamic fluctuations.

Another effect caused by hydrodynamic fluctuations
is the decorrelation of the flow angles
at different rapidities
which have a strong correlation
due to the initial-state fluctuations
from nucleon position distributions.
The decorrelations
can be caused by several mechanisms
\cite{Jia:2014ysa}
including hydrodynamic fluctuations.
In the right panel of Fig.~\ref{fig:vnExE}
we see the decorrelation by hydrodynamic fluctuations.
The flow angles are disturbed by the hydrodynamic fluctuations
which have short longitudinal correlations due to the FDR.

\section{Summary and discussions}
Hydrodynamic fluctuations have
a non-negligible amount of effects on observables of flow harmonics
compared to that of the initial-state fluctuations.
In central collisions
hydrodynamic fluctuations
increase the flow coefficients.
The increase is larger in higher order of harmonics
as expected from the FDR.
In non-central collisions with a large geometrical origin of $v_2$,
hydrodynamic fluctuations do not change
$v_2$ of the event-plane method
which roughly corresponds to the average of event-by-event $v_2$.
Nevertheless the distribution of the event-by-event $v_2$
is broadened by the fluctuations.
Also, hydrodynamic fluctuations
contribute to the decorrelation of the flow angles
in the longitudinal direction.

The effects of hydrodynamic fluctuations are too large
with the current parameter set of the shear viscosity, $\eta/s$,
and the cutoff scale of hydrodynamic fluctuations, $\sigma_\eta$ and $\sigma_\perp$,
We need to fix the realistic values of those parameters
by scanning the parameters in event-by-event calculations.
Also there is room to improve the model
by taking into account
the corrections of various quantities
by the cutoff scale of hydrodynamic fluctuations.
For example, transport coefficients should be renormalized
for each cutoff scale of hydrodynamic fluctuations
\cite{Kovtun:2012rj}.
The equation of state,
the phasespace distribution of hadrons
on the switching hypersurface
and even the initial entropy density
could be subject to change due to renormalization.

This work was supported by JSPS KAKENHI Grant Numbers 12J08554 (K.M.) and 25400269 (T.H.).





\bibliographystyle{elsarticle-num}
\bibliography{kmurase}

\begin{thebibliography}{10}
\expandafter\ifx\csname url\endcsname\relax
  \def\url#1{\texttt{#1}}\fi
\expandafter\ifx\csname urlprefix\endcsname\relax\def\urlprefix{URL }\fi
\expandafter\ifx\csname href\endcsname\relax
  \def\href#1#2{#2} \def\path#1{#1}\fi

\bibitem{Kapusta:2011gt}
J.~I. Kapusta, B.~Muller, M.~Stephanov, Phys. Rev. C85 (2012) 054906.
\newblock \href {http://arxiv.org/abs/1112.6405} {\path{arXiv:1112.6405}},
  \href {http://dx.doi.org/10.1103/PhysRevC.85.054906}
  {\path{doi:10.1103/PhysRevC.85.054906}}.

\bibitem{Landau:1980mil}
L.~D. Landau, E.~M. Lifshitz, {Fluid Mechanics}, Vol.~6 of Course of
  Theoretical Physics, Pergamon Press, New York, 1959.

\bibitem{Hirano:2012kj}
T.~Hirano, P.~Huovinen, K.~Murase, Y.~Nara, Prog. Part. Nucl. Phys. 70 (2013)
  108--158.
\newblock \href {http://arxiv.org/abs/1204.5814} {\path{arXiv:1204.5814}},
  \href {http://dx.doi.org/10.1016/j.ppnp.2013.02.002}
  {\path{doi:10.1016/j.ppnp.2013.02.002}}.

\bibitem{Drescher:2007ax}
H.-J. Drescher, Y.~Nara, Phys. Rev. C76 (2007) 041903.
\newblock \href {http://arxiv.org/abs/0707.0249} {\path{arXiv:0707.0249}},
  \href {http://dx.doi.org/10.1103/PhysRevC.76.041903}
  {\path{doi:10.1103/PhysRevC.76.041903}}.

\bibitem{Cooper:1974mv}
F.~Cooper, G.~Frye, Phys. Rev. D10 (1974) 186.
\newblock \href {http://dx.doi.org/10.1103/PhysRevD.10.186}
  {\path{doi:10.1103/PhysRevD.10.186}}.

\bibitem{Teaney:2003kp}
D.~Teaney, Phys. Rev. C68 (2003) 034913.
\newblock \href {http://arxiv.org/abs/nucl-th/0301099}
  {\path{arXiv:nucl-th/0301099}}, \href
  {http://dx.doi.org/10.1103/PhysRevC.68.034913}
  {\path{doi:10.1103/PhysRevC.68.034913}}.

\bibitem{Nara:1999dz}
Y.~Nara, N.~Otuka, A.~Ohnishi, K.~Niita, S.~Chiba, Phys. Rev. C61 (2000)
  024901.
\newblock \href {http://arxiv.org/abs/nucl-th/9904059}
  {\path{arXiv:nucl-th/9904059}}, \href
  {http://dx.doi.org/10.1103/PhysRevC.61.024901}
  {\path{doi:10.1103/PhysRevC.61.024901}}.

\bibitem{Huovinen:2009yb}
P.~Huovinen, P.~Petreczky, Nucl. Phys. A837 (2010) 26--53.
\newblock \href {http://arxiv.org/abs/0912.2541} {\path{arXiv:0912.2541}},
  \href {http://dx.doi.org/10.1016/j.nuclphysa.2010.02.015}
  {\path{doi:10.1016/j.nuclphysa.2010.02.015}}.

\bibitem{Kovtun:2004de}
P.~Kovtun, D.~T. Son, A.~O. Starinets, Phys. Rev. Lett. 94 (2005) 111601.
\newblock \href {http://arxiv.org/abs/hep-th/0405231}
  {\path{arXiv:hep-th/0405231}}, \href
  {http://dx.doi.org/10.1103/PhysRevLett.94.111601}
  {\path{doi:10.1103/PhysRevLett.94.111601}}.

\bibitem{Song:2009gc}
H.~Song, Ph.D. thesis, Ohio State U. (2009).
\newblock \href {http://arxiv.org/abs/0908.3656} {\path{arXiv:0908.3656}},
  \href{http://inspirehep.net/record/829461/files/arXiv:0908.3656.pdf}{[link]}.
\newline\urlprefix\url{http://inspirehep.net/record/829461/files/arXiv:0908.3656.pdf}

\bibitem{Baier:2007ix}
R.~Baier, P.~Romatschke, D.~T. Son, A.~O. Starinets, M.~A. Stephanov, JHEP 04
  (2008) 100.
\newblock \href {http://arxiv.org/abs/0712.2451} {\path{arXiv:0712.2451}},
  \href {http://dx.doi.org/10.1088/1126-6708/2008/04/100}
  {\path{doi:10.1088/1126-6708/2008/04/100}}.

\bibitem{Murase:2013tma}
K.~Murase, T.~Hirano, {Unpublished results} (2013).
\newblock \href {http://arxiv.org/abs/1304.3243} {\path{arXiv:1304.3243}}.

\bibitem{Adler:2003cb}
S.~S. Adler, et~al., Phys. Rev. C69 (2004) 034909.
\newblock \href {http://arxiv.org/abs/nucl-ex/0307022}
  {\path{arXiv:nucl-ex/0307022}}, \href
  {http://dx.doi.org/10.1103/PhysRevC.69.034909}
  {\path{doi:10.1103/PhysRevC.69.034909}}.

\bibitem{Khachatryan:2015oea}
V.~Khachatryan, et~al., Phys. Rev. C92~(3) (2015) 034911.
\newblock \href {http://arxiv.org/abs/1503.01692} {\path{arXiv:1503.01692}},
  \href {http://dx.doi.org/10.1103/PhysRevC.92.034911}
  {\path{doi:10.1103/PhysRevC.92.034911}}.

\bibitem{Jia:2014ysa}
J.~Jia, P.~Huo, Phys. Rev. C90~(3) (2014) 034915.
\newblock \href {http://arxiv.org/abs/1403.6077} {\path{arXiv:1403.6077}},
  \href {http://dx.doi.org/10.1103/PhysRevC.90.034915}
  {\path{doi:10.1103/PhysRevC.90.034915}}.

\bibitem{Kovtun:2012rj}
P.~Kovtun, J. Phys. A45 (2012) 473001.
\newblock \href {http://arxiv.org/abs/1205.5040} {\path{arXiv:1205.5040}},
  \href {http://dx.doi.org/10.1088/1751-8113/45/47/473001}
  {\path{doi:10.1088/1751-8113/45/47/473001}}.

\end{thebibliography}







\end{document}